\documentclass[prd,twocolumn,superscriptaddress,showpacs,nofootinbib,preprintnumbers]{revtex4}

\usepackage{amsmath}
\usepackage{amsfonts}
\usepackage{graphicx}
\usepackage{dcolumn}
\usepackage{hyperref}

\newcommand{\be}{\begin{equation}}
\newcommand{\ee}{\end{equation}}
\newcommand{\bea}{\begin{eqnarray}}
\newcommand{\eea}{\end{eqnarray}}
\newcommand{\beaa}{\begin{eqnarray*}}
\newcommand{\eeaa}{\end{eqnarray*}}

\newcommand{\nn}{\nonumber \\}

\newcommand{\e}{{\rm e}}

\begin{document}

\tolerance=5000

\title{The oscillating dark energy: future singularity and coincidence
problem}

\author{Shin'ichi Nojiri}
\thanks{Electronic address: snojiri@yukawa.kyoto-u.ac.jp, \\ nojiri@cc.nda.ac.jp. \\
Address from April: Dept. of Phys., Nagoya Univ., Nagoya 464-8602, Japan.},
\affiliation{Department of Applied Physics, National Defence Academy,
Hashirimizu Yokosuka 239-8686, Japan}
\author{Sergei D. Odintsov}
\thanks{Electronic address: odintsov@ieec.uab.es also at TSPU, Tomsk}
\affiliation{Instituci\`{o} Catalana de Recerca i Estudis Avan\c{c}ats
(ICREA) and Institut de Ciencies de l'Espai (IEEC-CSIC),
Campus UAB, Facultat Ciencies, Torre C5-Par-2a pl, E-08193 Bellaterra
(Barcelona), Spain}

\begin{abstract}

We consider the oscillating dark energy with periodic equation of state
in two equivalent formulations: ideal fluid or scalar-tensor theory.
It is shown that such dark energy suggests the natural way for the
unification of early-time inflation with late-time acceleration.
We demonstrate how it describes the transition from deceleration to
acceleration or from
non-phantom to phantom era and how it solves the coincidence problem.
The occurence of finite-time future singularity for the oscillating
(phantom) universe is also investigated.

\end{abstract}


\maketitle

\section{Introduction.}

It became clear recently that current universe is expanding with
acceleration which is caused by dark energy (for a
recent review, see\cite{padmanabhan}). However, the observational data
(for a recent review, see \cite{obs}) are far from being complete.
It is not even known what is the current value of the dark energy
 effective equation of state
(EoS) parameter  $w$ which lies close to $-1$: it could be equal to $-1$
(standard $\Lambda$CDM cosmology), a little bit upper than $-1$ (the
quintessence dark energy \cite{padmanabhan}) or less than $-1$
(phantom dark energy, see \cite{phantom} for very incomplete list of
related papers).
Even less is known about the evolution of this EoS parameter, the origin
of dark energy and why currently the energy-density of dark energy is
approximately equal to the one of dust matter (coincidence problem).
As a result, there are numerous proposals for dark energy which so far
does not exist as consistent theory.

The oscillating dark energy has been proposed in
refs.\cite{osc1,osc2,osc3}. It has been shown there that such models very
naturally resolve the coincidence problem due to periods of acceleration
and that they may be consistent with observations. They also serve as very
good candidates for unification of the early time inflation with late time
acceleration. It was argued that oscillated dark energy does not lead to
Big Rip singularity.

In the present paper the late-time cosmological consequences of dark
energy with time-dependent, periodic EoS  are further described.
We work in two (mathematically equivalent) formulations: ideal fluid EoS
description
or scalar-tensor theory with some specific potential. The general method
 to go from one formulation to another is proposed.
It is demonstrated for number of examples that such oscillating dark
energy may describe the
transition from deceleration to acceleration, or from non-phantom to
phantom era (if current dark energy is of phantom type).
In the last case, the classical universe may end up in the finite-time,
future singularity (any of four known types of Big Rip may occur as is
shown explicitly). The fact that oscillating dark energy universe
may end up in the future singularity was not discussed before.
It is proved that the oscillating dark energy may be the key
for resolution of the coincidence problem in accordance with earlier
findings of refs.\cite{osc1,osc2}. Finally, we show that the
unification of early-time
inflation with late-time acceleration is very natural in such oscillating
universe which could be embedded in the cyclic (expanding/contracting)
universe.

\section{Oscillating equation of state of the universe.}

Let us consider the universe filled with the ideal fluid (dark energy)
where
the equation of state (EoS) parameter $w$ depends on time $t$:
\be
\label{T1}
p=w(t)\rho\ .
\ee
Recent observational data (for review, see \cite{obs}) admit the
possibility of time-dependent $w$ which lies currently close to $-1$.

 The conservation of the energy
\be
\label{T2}
\dot\rho + 3H\left(p+\rho\right)=0\ ,
\ee
and  the FRW equation
\be
\label{T3}
\frac{3}{\kappa^2}H^2 = \rho\ ,
\ee
give
\be
\label{T4}
\dot\rho + \kappa\sqrt{3}\left(1+w(t)\right)\rho^{3/2}=0\ ,
\ee
which can be integrated as
\be
\label{T5}
\rho=\frac{4}{3\kappa^2\left(\int dt \left(1+w(t)\right)\right)^2}\ .
\ee
Using (\ref{T4}), the Hubble rate may be found
\be
\label{T6}
H=h(t)\equiv \frac{2}{3\int dt \left(1+w(t)\right)}\ .
\ee
When $w$ is a constant (not equal to phantom divide value), the standard
expression is recovered
\be
\label{T7}
H=\frac{2}{3\left(1+w\right)\left(t-t_s\right)}\ .
\ee
If $\int dt \left(1+w(t)\right)=0$, $H$ diverges,
which corresponds to the Big Rip type singularity,
which occurs when $t=t_s$.
 In case $w>-1$, for the expanding universe
where $H>0$, the cosmological time $t$ should be restricted to be $t>t_s$.

The situation where $w(t)$ is time-dependent, periodic function may be of
physical interest. Indeed, this can explain easily why the
effective value of $w$ is different at different epochs, being currently
close to $-1$ (if the period of such function is comparable with the
universe age).
An illustrative example is given by
\be
\label{T8}
w= -1 + w_0 \cos \omega t\ .
\ee
Here, $w_0,\omega>0$.
(Such EoS has been considered in ref.\cite{osc2} where it was shown
that it may be consistent with observations for some values of parameters.)
Then Eq.(\ref{T8}) gives
\be
\label{T9}
H = \frac{2\omega}{3\left(w_1 + w_0 \sin \omega t\right)}\ .
\ee
Here $w_1$ is a constant of the integration. When $|w_1|<w_0$, the denominator can vanish,
which corresponds to the Big Rip singularity. When $|w_1|>w_0$, there does
not occur such
a singularity.
Since
\be
\label{T10}
\dot H = - \frac{2\omega^2 w_0 \cos \omega t}{3\left(w_1 + w_0 \sin \omega t\right)^2}\ ,
\ee
when $w_0 \cos \omega t<0$ $\left(w_0 \cos \omega t>0\right)$,
the universe lives in phantom
(non-phantom) phase where $\dot H>0$ $\left(\dot H<0\right)$.
In this case, the energy density $\rho$ is given by
\be
\label{T10a}
\rho(t)=\frac{4\omega^2}{3\kappa^2\left(w_1 + w_0 \sin \omega t\right)^2}\ ,
\ee
which also oscillates. Hence,
$\dot\rho>0$ in phantom phase and $\dot\rho<0$ in non-phantom phase.
Hence, kind of energy transfer occurs in such oscillating dark energy
universe: the energy-density grows in phantom
phase but decreases in non-phantom era.
This may suggest the qualitative explanation of the fact of the growth of
phantom energy.
Moreover, in such oscillating universe the unification of late-time
(phantom or quintessence) acceleration with early-time, phantom inflation
\cite{inflation} naturally occurs: indeed, the (phantom or quintessence)
acceleration epoch occurs periodically due to
 the periodic behaviour of the universe EoS.

 Since the pressure is given by
\be
\label{T11c}
p=-\frac{1}{\kappa^2}\left(2\dot H + 3H^2\right)\ ,
\ee
from the second FRW equation, we find
\be
\label{T11d}
p(t)=\frac{4\omega^2\left(w_0\cos\omega t - 1\right)}
{3\kappa^2\left(w_1 + w_0 \sin \omega t\right)^2}\ ,
\ee
Then as long as $w_1 + w_0 \sin \omega t>0$, that is,
the universe is expanding $\left(H>0\right)$,
we find the following EoS for dark energy ideal fluid:
\be
\label{T13}
w_0^2 = \left( w_1 - \frac{2\omega}{\kappa\sqrt{3\rho}} \right)^2
+ \left(1 + \frac{p}{\rho}\right)^2 \ .
\ee

Another interesting example considered in ref.\cite{osc1} is
\bea
\label{TT1}
&& w=-1 + \alpha\left( 1 + \beta \cos \omega t\right)\ ,\nn
&& \alpha>0\ ,\quad 1>\beta>0\ .
\eea
Here $w>-1$ and the universe is not in phantom phase.
Hence, this model corresponds to oscillating dark
energy where the unification of early-time quintessence inflation with
late-time quintessence acceleration occurs.
Moreover, if $\alpha\left(1+\beta\right)>1$, $w$ crosses $w=0$ and
we find
\be
\label{TT3}
H=\frac{2\omega}{3\alpha\left\{\omega(t-t_0) + \beta\sin\omega t\right\}}\ ,
\ee
Here $t_0$ is a constant of the integration.
In this case, the energy density $\rho$ is given by
\be
\label{TT3a}
\rho=\frac{4\omega^2}{3\kappa^2\alpha^2\left\{\omega(t-t_0) + \beta\sin\omega t\right\}^2}\ ,
\ee
which is monotonically decreasing, oscillating function.
Since
\be
\label{TT3b}
p=\frac{4\omega^2\left\{(\alpha - 1) + \alpha\beta \cos\omega t\right\}}
{3\kappa^2\alpha^2\left\{\omega(t-t_0) + \beta\sin\omega t\right\}^2}\ ,
\ee
we find the following EoS for ideal fluid:
\bea
\label{TT3c}
0&=&\frac{p+(1-\alpha)\rho}{\alpha\beta \rho} - \cos\Bigl(\omega t_0
+ \frac{\alpha\kappa}{2\omega}\sqrt{3\rho} \nn
&& + \frac{1}{\alpha\rho}\left\{\left(\alpha^2\beta^2 - \alpha^2
+ 2\alpha - 1\right)\rho^2\right.\nn
&& \left. - p^2 - 2(1-\alpha)p\rho\right\}^\frac{1}{2}\Bigr)\ .
\eea

In a more realistic situation, universe is filled with matter and dark
energy.
The matter, in general, interacts with the dark energy. In such a case,
 the total energy density  $\rho_{\rm tot}$ consists of
the contributions from the dark energy and the matter:
$\rho_{\rm tot}=\rho + \rho_m$. If we define, however, the matter energy
density $\rho_m$ properly, we can also {\it define} the matter pressure $p_m$
and the dark energy pressure $p$ by
\be
\label{Sep1}
p_m\equiv -\rho_m + \frac{\dot \rho}{3H}\ , \quad
p\equiv p_{\rm tot} - p_m\ .
\ee
Here $p_{\rm tot}$ is the total pressure.
Hence, the  matter and dark energy
satisfy the energy conservation laws separately,
\be
\label{Sep2}
\dot \rho_m + 3H\left(\rho_m + p_m\right)=0\ ,\quad
\dot \rho + 3H\left(\rho + p\right)=0\ .
\ee

We now consider the case that universe is filled
with the oscillating dark energy and the matter ideal fluid with a
constant
EoS parameter $w_m$: $p_m=w_m\rho_m$. Since $w_m$ is a constant, one finds
\be
\label{T19}
\rho_m=\rho_0 a^{-3\left(1+w_m\right)}\ .
\ee
Here $\rho_0$ is a constant.
 FRW equation leads to
\be
\label{T20}
\rho=\frac{3}{\kappa^2}H^2 - \rho_0 a^{-3\left(1+w_m\right)}\ .
\ee
Using the conservation law (\ref{T2}), we find
\bea
\label{T21}
&& -3\left(1+w(t)\right)=\frac{d}{dt}\ln\rho \nn
&& = \frac{\frac{6}{\kappa^2}H\dot H + 3\left(1+w_m\right) H \rho_0 a^{-3\left(1+w_m\right)}}
{\frac{3}{\kappa^2}H^2 - \rho_0 a^{-3\left(1+w_m\right)}}\ .
\eea
That is,
\be
\label{T22}
1+w(t) = - \frac{\frac{2}{\kappa^2}\dot H + \left(1+w_m\right)  \rho_0 a^{-3\left(1+w_m\right)}}
{\frac{3}{\kappa^2}H^2 - \rho_0 a^{-3\left(1+w_m\right)}}\ .
\ee
Note that the denominator of the r.h.s. in (\ref{T22}) is always positive
since the denominator is nothing
but energy-density.

Let us consider the  transition from non-phantom  to the phantom era,
where $\dot H=0$.
Then  $1+w(t)<0$ as long as $w_m>-1$. Thus, in order that the transition
occurs, $w(t)$ should become
less than $-1$ before the transition.

As an example, we consider the scale factor
\be
\label{T29}
a(t)=a_0\left(\frac{t}{t_s - t}\right)^{h_0}\ ,
\ee
with constants $a_0$, $t_s$, and $h_0>0$. Eq.(\ref{T29}) gives,
\bea
\label{T30}
H &=& \frac{h_0t_s}{t\left(t_s - t \right)}\ ,\nn
\frac{\ddot a}{a}&=&\frac{h_0t_s\left(2t - \left(1 - h_0\right)t_s\right)}{t^2\left(t_s - t\right)^2}\ .
\eea
The transition from non-phantom to phantom era occurs at $t=t_s/2$,
and if $h_0<1$, the transition from decceleration to acceleration epoch
occurs at $t=\frac{t_s\left(1-h_0\right)}{2}$.
Using (\ref{T22}),  $w(t)$ corresponding to (\ref{T29}) is given by
\bea
\label{T33}
&& w(t) = -1 \nn
&& - \left\{\frac{2h_0 t_s \left(2t - t_s\right)}{\kappa^2t^2\left(t_s - t\right)^2} \right. \nn
&& \left. + \left(1+w_m\right)  \rho_0 a_0^{-3\left(1+w_m\right)}
\left(\frac{t}{t_s - t}\right)^{-3\left(1+w_m\right)h_0}\right\} \nn
&& \times \left\{\frac{3h_0^2t_s^2}{\kappa^2t^2\left(t_s - t \right)^2} \right. \nn
&& \left. - \rho_0 a_0^{-3\left(1+w_m\right)}\left(\frac{t}{t_s - t}\right)^{-3\left(1+w_m\right)h_0}\right\}^{-1}\ .
\eea
Now the energy-density is given by
\bea
\label{T33b}
\rho(t)&=&\frac{3h_0^2t_s^2}{\kappa^2 t^2\left(t_s - t \right)^2} \nn
&& - \rho_0a_0^{-3\left(1+w_m\right)} \left(\frac{t}{t_s - t}\right)^{-3h_0\left(1+w_m\right)}\ .
\eea
Since the second term in (\ref{T33b}) is negative, $\rho(t)$ is not always positive.
The negative energy density is, of course, not physical.
When $t\to 0$, if $3h_0\left(1+w_m\right)>2$, the second term dominates and the energy density
becomes negative. On the other hand,
when $t\to t_s$, if $3h_0\left(1+w_m\right)<-2$, one finds
that $\rho$ becomes negative, again.
Hence, physically allowed region of the parameters could be
\be
\label{T52}
-2\leq 3h_0\left(1+w_m\right)\leq 2\ .
\ee
For simplicity, we consider the case $3h_0 \left(1 + w_m\right)=2$.
Then the expression (\ref{T33b}) reduces to
\bea
\label{T33g}
\rho(t)&=&\frac{1}{t^2}Q(t)\ ,\nn
Q(t)&\equiv& \frac{3h_0^2 t_s^2}{\kappa^2 \left(t-t_s\right)^2}
 - \rho_0 a_0^{-\frac{2}{h_0}}\left(t_s - t\right)^2\ .
\eea
Since $Q(t)$ is monotonicaly increasing function as long as $t<t_s$, if
\be
\label{T33h}
Q(0)= \frac{3h_0^2 }{\kappa^2} - \rho_0 a_0^{-\frac{2}{h_0}}t_s^2\geq 0\ ,
\ee
$\rho(t)$ is always positive.

 From the second FRW equation, the pressure may be found:
\be
\label{TT33c}
p=-\frac{1}{\kappa^2}\left(2\dot H + 3 H^2\right) - w_m \rho_0 a^{-3\left(1+w_m\right)}\ .
\ee
Hence,
\bea
\label{T33d}
&& p(t)=- \frac{2h_0t_s\left(2t - t_s\right) + 3h_0^2t_s^2}{\kappa^2 t^2\left(t_s - t \right)^2} \nn
&& - w_m \rho_0a_0^{-3\left(1+w_m\right)} \left(\frac{t}{t_s - t}\right)^{-3h_0\left(1+w_m\right)}\ .
\eea
At the phantom-non-phantom transition point $t=t_s/2$, one gets
\bea
\label{T33e}
&& \rho(t) + p(t) \nn
&& = - \left(1+w_m\right) \rho_0a_0^{-3\left(1+w_m\right)} \nn
&& \times \left(\frac{t}{t_s - t}\right)^{-3h_0\left(1+w_m\right)}\ ,
\eea
which is  negative, corresponding to $w(t)<-1$ there.

As the next example, the following scale factor may be considered:
\be
\label{T34}
a=a_0\e^{h_0 t + \frac{h_1}{\omega} \sin \omega t}\ ,
\ee
with constants $h_0>h_1>0$ and $\omega>0$.
Since
\be
\label{T35}
H=h_0 + h_1 \cos \omega t\ .
\ee
Here, the universe lives in non-phantom (phantom) era when
$2\pi n < \omega t < 2\pi n + \pi$ $\left(2\pi n - \pi < \omega t < 2\pi n \right) $,
where $n$ is an integer.
The corresponding, time-dependent EoS parameter $w(t)$ is found to be
\bea
\label{T36}
&& w(t)= -1 \nn
&& - \Bigl\{ - \frac{2h_1}{\kappa^2} \sin \omega t \nn
&& + \left(1+w_m\right) \rho_0 a_0^{-3\left(1+w_m\right)}
\e^{-3\left(1+w_m\right)\left(h_0 t + \frac{h_1}{\omega} \sin \omega t\right)}\Bigr\} \nn
&& \times \Bigl\{ \frac{3}{\kappa^2} \left( h_0 + h_1 \cos \omega t \right)^2 \nn
&& - \rho_0 a_0^{-3\left(1+w_m\right)}
\e^{-3\left(1+w_m\right)\left(h_0 t + \frac{h_1}{\omega} \sin \omega t\right)} \Bigr\}^{-1} \ .
\eea
Now the energy density $\rho$ and pressure $p$ are
\bea
\label{T36b}
&& \rho(t)=\frac{3}{\kappa^2}\left(h_0 + h_1 \cos \omega t\right)^2 \nn
&& \quad - \rho_0a_0^{-3\left(1+w_m\right)}
\e^{-3\left(1 + w_m \right)\left(h_0 t + \frac{h_1}{\omega} \sin \omega t\right)} \ , \nn
&& p(t)=\frac{1}{\kappa^2}\left\{ 2h_1\omega \sin \omega t - 3\left(h_0 + h_1 \cos \omega t\right)^2\right\} \nn
&& \quad - w_m\rho_0a_0^{-3\left(1+w_m\right)}
\e^{-3\left(1 + w_m \right)\left(h_0 t + \frac{h_1}{\omega} \sin \omega t\right)} \ .
\eea
The  $\rho(t)$  (\ref{T36b}) becomes negative
 for large negative $t$. If we restrict the parameters to satisfy
\be
\label{T36c}
\frac{3}{\kappa^2}\left(h_0 - h_1 \right)^2 - \rho_0a_0^{-3\left(1+w_m\right)}\geq 0\ ,
\ee
it follows $\rho(t)\geq 0$ when $t\geq 0$.

It is interesting to understand if the oscillating dark energy
may bring the universe evolution to Big Rip singularity.
(One should not forget that Big Rip is typically classical effect which
seems to disappear with proper account of quantum corrections
\cite{quantum,final,
NOT}.) The classification of future, finite-time singularities
is given as\cite{NOT}:
\begin{itemize}
\item  Type I (``Big Rip'') : For $t \to t_s$, $a \to \infty$,
$\rho \to \infty$ and $|p| \to \infty$
\item  Type II (``sudden'') : For $t \to t_s$, $a \to a_s$,
$\rho \to \rho_s$ or $0$ and $|p| \to \infty$
\item  Type III : For $t \to t_s$, $a \to a_s$,
$\rho \to \infty$ and $|p| \to \infty$
\item  Type IV : For $t \to t_s$, $a \to a_s$,
$\rho \to 0$, $|p| \to 0$ and higher derivatives of $H$ diverge.
This also includes the case when $\rho$ ($p$) or both of them
tend to some finite values while higher derivatives of $H$ diverge.
\end{itemize}
Here $t_s$, $a_s$ and $\rho_s$ are constants with $a_s\neq 0$.

Let assume
\be
\label{T14}
H \sim h_0 \left(t_s - t\right)^{-\alpha}\ ,\quad \left(h_0>0\right)\ ,
\ee
near the singularity $t\sim t_s$. Then one gets
\bea
\label{T15}
\dot H &\sim & \left\{\begin{array}{ll}
h_0\left(t_s - t\right)^{-\alpha-1} & (\alpha \neq 0) \\
0 & (\alpha = 0 )
\end{array} \right. \nn
\ln a &\propto& \left\{\begin{array}{ll}
h_0\left(t_s - t\right)^{1-\alpha} & (\alpha \neq 1) \\
h_0\ln \left(t_s - t\right) & (\alpha = 1 )
\end{array} \right. \ .
\eea
Since $\rho \propto \sqrt{H}$ and $p \propto 3H^2 + 2\dot H$, in terms of $\alpha$, the above
four types of the singularities could be classified as

\noindent
Type I: $\alpha \geq 1$, Type II: $0>\alpha>-1$, Type III: $1>\alpha>0$,
Type IV: $\alpha<-1$ but $\alpha$ is not an integer (for account of
Hubble rate dependent terms in inhomogeneous EoS, see \cite{brevik2}).

Without the matter, in terms of $w(t)$,
by comparing (\ref{T14}) with (\ref{T6}), we find
\be
\label{T16}
1 + w(t) \sim \left(t_s - t\right)^{\alpha - 1}\ .
\ee
Thus, all four types of future singularities are realized here for
above values of parameter $\alpha$.

It is interesting to study what happens with the classical future
singularity in the presence  of matter. For this purpose,
we assume Eq.(\ref{T14}). Then from (\ref{T22}), we find
\bea
\label{T23}
&& 1+w(t) \nn
&& \sim - \Bigl(-\frac{2\alpha}{\kappa^2}h_0\left(t_s - t\right)^{-\alpha -1} \nn
&& + 3\left(1+w_m\right)\rho_0 a_0
\e^{-\frac{h_0\left(1 + w_m\right)}{1-\alpha} \left(t_s - t\right)^{-\alpha +1}}\Bigr) \nn
&& \times \Bigl(\frac{3}{\kappa^2}h_0^2\left(t_s - t\right)^{-2\alpha} \nn
&& - \rho_0 a_0\e^{-\frac{h_0\left(1 + w_m\right)}{1-\alpha} \left(t_s - t\right)^{-\alpha +1}}\Bigr)^{-1}\ ,
\eea
when $\alpha\neq 1$ and for $\alpha=1$
\bea
\label{T24}
&& 1+w(t) \nn
&& \sim - \Bigl(-\frac{2\alpha}{\kappa^2}h_0\left(t_s - t\right)^{-2} \nn
&& + 3\left(1+w_m\right)\rho_0 a_0 \left(t_s - t\right)^{-h_0\left(1 + w_m\right)}\Bigr) \nn
&& \times \Bigl(\frac{3}{\kappa^2}h_0^2\left(t_s - t\right)^{-2} \nn
&& - \rho_0 a_0\left(t_s - t\right)^{-h_0\left(1 + w_m\right)}\Bigr)^{-1}\ .
\eea

In case $\alpha>1$, which corresponds to Type I singularity,
one obtains
\be
\label{T25}
1+w(t)\sim
\frac{2\alpha}{3h_0}\left(t_s - t\right)^{\alpha - 1} \to 0\ .
\ee
if $1+w_m<0$. The case $1+w_m>0$ is excluded since the denominator of the
r.h.s.
in (\ref{T22}) becomes negative.

In case $\alpha=1$, which also corresponds to Type I singularity,
\be
\label{T26}
1+w(t)\sim \left\{
\begin{array}{ll}
\frac{-\frac{2\alpha}{\kappa^2}h_0 + 3\left( 1 + w_m \right)\rho_0 a_0}
{\frac{3}{\kappa^2}h_0^2  - \rho_0 a_0} & \mbox{if}\ h_0\left(1+ w_m \right)=2 \\
- \frac{2\alpha}{3h_0} & \mbox{if}\ h_0\left(1+w_m \right)=2 \\
\end{array}\right. \ .
\ee
The case $h_0\left(1+w_m \right)>2$ is also prohibited since the denominator of
the r.h.s. in (\ref{T22}) becomes negative.

In case $0<\alpha<1$, which corresponds to Type III singularity, we find
\be
\label{T27}
1+w(t)\sim
\frac{2\alpha}{3h_0}\left(t_s - t\right)^{\alpha - 1} \to \infty\ ,
\ee
and in case $\alpha=0$,
\be
\label{T28}
1+w(t)\sim
\frac{3\left(1+w_m\right)\rho_0 a_0}{\frac{3}{\kappa^2}h_0^2 - \rho_0 a_0}\ .
\ee

The case $-1<\alpha<0$, which corresponds to Type II singularity, and
the case $\alpha<-1$, which corresponds to Type IV, are excluded since the
denominator of
the r.h.s. in (\ref{T22}) becomes negative.
It is interesting that the Type II and IV singularities are excluded
 with the account of matter, while  they occur without matter.
Thus, we demonstrated that within oscillating dark energy universe, all
four types
of future singularities may occur on the classical level. This is our main
qualitative result: the oscillating dark energy of general form does not
prevent the universe to end up in the future singularity of any from four
known types.

\section{Scalar-tensor description and coincidence problem}

In this section we will present the oscillating dark energy with
time-dependent, explicit EoS in an equivalent, scalar-tensor
description following to the method developed in ref.\cite{eq}.
 (Note that such method may be applied to include also the time-dependent
bulk viscosity \cite{eq1} introduced for dark energy description in
ref.\cite{brevik}.)
Let us
start with the following action
\bea
\label{k1}
S&=&\int d^4 x \sqrt{-g}\Bigl\{\frac{1}{2\kappa^2}R
 - \frac{1}{2}\Omega(\phi)\partial_\mu \phi \partial^\mu \phi \nn
&& - V(\phi)\Bigr\} \ , \nn
\Omega(\phi) &=&- \frac{2}{\kappa^2}f'(\phi)\ ,\nn
V(\phi)&=&\frac{1}{\kappa^2}\left(3f(\phi)^2 + f'(\phi)\right)\ ,
\eea
with an adequate function $f(\phi)$.  The following solution of
FRW equation exists\cite{CNO}:
\be
\label{k7}
\phi=t\ ,\quad H=f(t)\ .
\ee
Then by comparing (\ref{k7}) with (\ref{T6}), one may identify
\be
\label{T11}
f(t)=h(t)\ .
\ee
For the action (\ref{k1}),  the explicit EoS is given by:
\be
\label{SN1}
p=-\rho -
\frac{2}{\kappa^2}f'\left(f^{-1}\left(\kappa\sqrt{\frac{\rho}{3}}\right)\right)\ ,
\ee
Thus, we demonstrated how an arbitrary oscilating dark energy may be
rewritten explicitly in mathematically-equivalent scalar-tensor form.

Hence, for Eq.(\ref{T8}),  the following equivalent description in
scalar-tensor form may be constructed:
\bea
\label{T12}
\Omega(\phi) &=& \frac{4\omega^2 w_0 \cos \omega \phi}
{3\kappa^2\left(w_1 + w_0 \sin \omega \phi\right)^2}\ ,\nn
V(\phi) &=& \frac{\omega^2 \left(4 - 2 w_0 \cos \omega \phi\right)}
{3\kappa^2\left(w_1 + w_0 \sin \omega \phi\right)^2}\ .
\eea
For EoS (\ref{TT1}), the equivalent scalar-tensor theory
description \cite{eq} leads
to
\bea
\label{TT4}
\Omega(\phi)&=&-\frac{4\omega^2 \left( 1 + \beta \cos \omega \phi\right)}
{3\alpha\kappa^2\left\{\Omega(\phi -t_0) + \beta\sin\omega \phi\right\}^2}\ ,\nn
V(\phi)&=&\frac{2\omega^2 \left( 2- \alpha - \alpha \beta \cos \omega \phi\right)}
{3\alpha^2\kappa^2\left\{\omega(\phi -t_0) + \beta\sin\omega \phi\right\}^2}\ .
\eea

Comparing (\ref{T14}) with (\ref{T11}), one also finds that in
scalar-tensor
theory (\ref{k1}), the four types of future singularities could be
realized by the choice
\be
\label{T17}
f(\phi) \sim f_0 \left(t_s - \phi\right)^{-\alpha} = f_0 \varphi^{-\alpha}\ ,
\ee
with positive constant $f_0$.
Here $\varphi\equiv t_s - \phi$. In this case, we obtain
\bea
\label{T18}
\Omega(\phi) &\sim& - \frac{2\alpha f_0}{\kappa^2}\left(t_s - \phi\right)^{-\alpha-1}\ ,\\
V(\phi)&=&\frac{1}{\kappa^2}\left(3f_0^2\left(t_s - \phi\right)^{-2\alpha}
+ \alpha f_0 \left(t_s - \phi\right)^{-\alpha-1}\right)\ ,\nonumber
\eea
One should bear in mind that even FRW equations are the same in EoS or
scalar-tensor descrition, the emerging universes may be different.
Indeed, the
number and stability of FRW solutions  may not coincide, the newton law
may be slightly different, etc.

As an extension of (\ref{k1}),
 the matter with constant EoS parameter $w=w_m$ may be added into the
action (\ref{k1}).
By using a single function $g(t)$, if we choose the scalar potentials in
the action (\ref{k1}) with matter as
\bea
\label{T37}
\Omega(\phi) &=&- \frac{2}{\kappa^2}g''(\phi) -
\frac{w_m + 1}{2}g_0 \e^{-3(1+w_m)g(\phi)}\ ,\nn
V(\phi) &=& \frac{1}{\kappa^2}\left(3g'(\phi)^2 + g''(\phi)\right) \nn
&& +\frac{w_m -1}{2}g_0 \e^{-3(1+w_m)g(\phi)} \ ,
\eea
the following solution of FRW equation may be constructed (see also
\cite{CNO}):
\bea
\label{T38}
&& \phi=t\ ,\quad H=g'(t)\ ,\nn
&& \left(a=a_0 \e^{g(t)}\ ,\quad a_0\equiv
\left(\frac{\rho_{m0}}{g_0}\right)^{\frac{1}{3(1+w_m)}}\right)
\eea
In the case of scale factor (\ref{T29})
\be
\label{T39}
g(\phi)=\ln \left(\frac{\phi}{t_s - \phi}\right)^{h_0}\ .
\ee
and in the case of (\ref{T34})
\be
\label{T40}
g(\phi) =\ln \left(h_0 \phi + \frac{h_1}{\omega} \sin \omega \phi\right)\ ,
\ee

Let us now consider a little bit different function
\be
\label{T41}
g(\phi)=\alpha \ln \left\{ \hat g_0 t + \frac{\hat g_1}{\omega}\cos\omega \phi\right\}\ .
\ee
Here $\alpha$, $\hat g_0$, $\hat g_1$, and $\omega$ are positive constants.
Then the Hubble rate is given by
\be
\label{T42}
H= \frac{\alpha \left(\hat g_0 - \hat g_1 \sin \omega t\right)}{\hat g_0 t + \frac{\hat g_1}{\omega}\cos\omega t}\ .
\ee
If $t>1/\omega$, the Hubble rate $H$ is always positive and if $\hat g_0$ is large enough
compared with $\hat g_1$ and $\omega$, we find $\dot H<0$.
Taking $\rho$ and $p$ in (\ref{T1}) as total energy density and pressure,
 $w(t)$ is
given by
\be
\label{T43}
w(t) = - 1 + \frac{2}{3\alpha} + \frac{2\hat g_1\omega\cos\omega t}{3\alpha
\left( \hat g_0 - \hat g_1 \sin \omega t\right)^2}\ .
\ee
When $\hat g_0\gg \hat g_1$, the expression (\ref{T43}) reduces to
\be
\label{T44}
w(t) = - 1 + \frac{2}{3\alpha} + \frac{2\hat g_1\omega\cos\omega t}{3\alpha \hat g_0^2}\ .
\ee
Then if
\be
\label{T45}
\frac{\hat g_1\omega}{\hat g_0^2}>\left| 1 - \alpha \right|\ ,
\ee
$w$ oscillates and crosses $w=-1/3$. Hence, the universe expiriences the
transition from the decceleration to the  acceleration epoch.

Having in mind the oscillating nature of dark energy under consideration,
it is expected \cite{osc1,osc2,osc3} that such type of dark energy may
naturally
solve the
coincidence problem (for recent discussion and list of refs., see
\cite{cai}). Let us confirm this in both formulations of the oscillating
dark energy under consideration.
We now separate $\rho_{\rm tot}$ into the contributions from the dark energy and matter:
$\rho_{\rm tot}=\rho + \rho_m$.
If the EoS parameter $w_m$ of the matter is constant as in (\ref{T19}), by using
the first FRW equation (\ref{T3}), one gets
\be
\label{T47}
r\equiv \frac{\rho}{\rho_m}= -1 + \frac{3H^2 a^{3\left(1+w_m\right)}}{\kappa^2 \rho_{m0}}\ .
\ee
For Eq.(\ref{T41}), it follows
\bea
\label{T48}
r &=& - 1 + \frac{3\alpha^2 a_0^{3(1+w_m)}}{\kappa^2\rho_{m0}}
\left(g_0 - g_1 \sin \omega t\right) \nn
&& \times \left(g_0 t + \frac{g_1}{\omega}\cos\omega t\right)^{3\left(1+w_m\right)\alpha - 2}\ .
\eea
With the choice $\alpha = 2/3\left(1+w_m\right)$, the Eq.(\ref{T48}) reduces to
\be
\label{T50}
r=- 1 + \frac{3\alpha^2 a_0^{3(1+w_m)}}{\kappa^2\rho_{m0}}
\left(g_0 - g_1 \sin \omega t\right)\ .
\ee
Then $r$ has a maximum and minimum as
\bea
\label{T51}
&& - 1 + \frac{3\alpha^2 a_0^{3(1+w_m)}}{\kappa^2\rho_{m0}}
\left(g_0 - g_1 \right) \leq r \nn
&& \leq
- 1 + \frac{3\alpha^2 a_0^{3(1+w_m)}}{\kappa^2\rho_{m0}}
\left(g_0 + g_1 \right)\ .
\eea
Hence, if $- 1 + \left(3\alpha^2 a_0^{3(1+w_m)}/\kappa^2\rho_{m0}\right)
\left(g_0 \pm g_1 \right)\sim {\cal O}(1)$, the ratio between the energy
densities of the dark energy and
matter could be always ${\cal O}(1)$. Hence, oscillating dark energy may
solve the coincidence problem.

In case of (\ref{T29}) or (\ref{T39}),  from (\ref{T33b}) one gets
\be
\label{T51b}
r = -1 + \frac{3h_0^2t_s^2 a_0^{3\left(1+w_m\right)}\left(t_s - t \right)^{-2-3h_0\left(1+w_m\right)}}
{\kappa^2 \rho_0 t^{2 - 3h_0\left(1+w_m\right)}} \ .
\ee
We now assume (\ref{T52}).

When $t\to t_s$, if $3h_0\left(1+w_m\right)\neq -2$, $r$ diverges but
if $3h_0\left(1+w_m\right)= -2$, $r$ is finite:
\be
\label{T53}
r \to r_0 \equiv -1 + \frac{3h_0^2 a_0^{3\left(1+w_m\right)}}{\kappa^2 \rho_0 t_s^2} \ ,
\ee
On the other hand, when $t=0$, if $3h_0\left(1+w_m\right)\neq 2$, $r$ diverges but
if $3h_0\left(1+w_m\right)=2$, $r$ is finite:
\be
\label{T53b}
r \to r_0 \ .
\ee
If we write $t=\gamma t_s$, when $3h_0\left(1+w_m\right)=2$, we find
\be
\label{T53c}
r \to r_l \equiv -1 + \frac{3h_0^2 a_0^{3\left(1+w_m\right)}}{\kappa^2 \rho_0 t_s^2
\left(1-\gamma\right)^4} \ ,
\ee
In the present universe, $0<\gamma<1$ and $\gamma\sim {\cal O}(1)$. In order that $r_0\sim {\cal O}(1)$,
$3h_0^2 a_0^{3\left(1+w_m\right)}/\kappa^2 \rho_0 t_s^2$ should be larger than
$1$ and ${\cal O}(1)$. Then we also find $r_l\sim {\cal O}(1)$. This tells that if the
ratio of the energy densities of the matter and the dark energy is ${\cal O}(1)$,
even in the present universe, the ratio is ${\cal O}(1)$,
which may solve the coincidence problem.

For the model in (\ref{T34}) or (\ref{T40}), by using (\ref{T36b}) one
obtains
\bea
\label{T54}
r &=& -1 + \frac{3a_0^{3\left(1+w_m\right)}}{\kappa^2\rho_0}\left(h_0 + h_1 \cos \omega t\right)^2 \nn
&& \times \e^{3\left(1 + w_m \right)\left(h_0 t + \frac{h_1}{\omega} \sin \omega t\right)} \ ,
\eea
which increases exponentially with time if $w_m>-1$.

Thus, we confirmed that oscillating dark energy may naturally solve the
coincidence problem in both (EoS or scalar-tensor) formulations which is
not clear from the beginning because the mathematical equivalence does not
mean the physical equivalence.

\section{Discussion}

In summary, we presented two equivalent formulations of oscillating dark
energy. It is demonstrated that such dark energy may describe
very naturally most relevant late-time cosmological phenomena:
 transition from
deceleration to acceleration era and transition from
non-phantom to phantom
epoch (if the universe currently enters such epoch). If the universe
currently lives in phantom era, it is demonstrated explicitly how the
oscillating dark energy may bring
the evolving universe to the one of four known types of finite-time,
future
singularity (on classical level). Moreover, oscillating dark energy may
suggest nice resolution of the coincidence problem as was suggested in
refs.\cite{osc1,osc2}. Finally, such dark
energy unifies in a clear and natural way the early-time inflation
and late-time acceleration whatever (quintessence, cosmological constant
or phantom) nature
they have.
However, in order to check if such unification is realistic one,
it is necessary to analyse the details of inflation, especially, the
preheating scenario. This lies beoynd the scopes of our paper because
some work in this direction (for phantom inflation) was already done in
\cite{inflation}.
It is also interesting that using method \cite{eq}
the oscillating dark energy may be presented as some special form of
modified
gravity (for recent review, see \cite{modified}).

One can go further and propose the following cosmological model:
\be
\label{T55}
H=h_0 \sin\omega t \left( 1 + h_1\sin m\omega t\right) \ .
\ee
Here $m$ is a positive integer, $h_0$, $h_1$, $\omega$ are positive
parameters and it is assumed that $m \gg 1$, $h_1 <1$.
When $2\pi n < \omega t < 2\pi n + \pi$ $\left(n\mbox{:integer}\right)$,
$H$ is positive
and universe is expanding but
when $2\pi n - \pi < \omega t < 2\pi n$  $\left(n\mbox{:integer}\right)$, $H$ is negative and
universe is contracting. Hence, the universe is oscillating with period
$2\pi/\omega$.
The conjecture could be that our cyclic universe (compare with
\cite{turok})
lives currently in the
expanding
era (repeating the acceleration/superacceleration circles each 14-15
billion years) while the transition from accelerating to contracting era
occurs with much bigger period of time.
Since it is assumed  $m\gg 1$,  adiabatically one obtains
\bea
\label{T57}
\dot H &=& h_0\omega\left\{ \cos\omega t \left( 1 + h_1\sin m\omega t\right) \right. \nn
&& \left. + h_1 m \sin\omega t \cos m\omega t\right\} \nn
&=& m \omega h_0 h_1 \sin\omega t \cos m\omega t\ .
\eea
Then in the expanding phase $\left(\sin\omega>0\right)$,
if $2\pi n - \pi/2 < \omega < 2\pi n + \pi/2$ $\left(n\mbox{:integer}\right)$,
we find $\dot H>0$, that is, the universe  enjoys the phantom era and
if $2\pi n + \pi/2 < \omega < 2\pi n + 3\pi/2$ $\left(n\mbox{:integer}\right)$,
$\dot H<0$, that is, the universe is in non-phantom era.
In the contracting phase $\left(\sin\omega<0\right)$,
if $2\pi n - \pi/2 < \omega < 2\pi n + \pi/2$ $\left(n\mbox{:integer}\right)$, $\dot H<0$
and if $2\pi n + \pi/2 < \omega < 2\pi n + 3\pi/2$ $\left(n\mbox{:integer}\right)$, $\dot H>0$.
Hence, the universe iterates phantom and non-phantom eras with the period
 $2\pi / m\omega$.
There is no problem to construct the EoS and/or scalar-tensor description
corresponding
to scale factor (\ref{T55}).
Nevertheless, the existing data cannot confirm/reject cyclic
cosmological models at high confidence level so one awaits the next
generation observational data which will be able to reconstruct the
realistic evolution
of the cosmological parameters.

\section*{Acknoweledgments}

The research by SDO has been supported in part by
the project FIS2005-01181 (MEC, Spain), by the project 06-01-00609 (RFBR,
Russia), by LRSS project N4489.2006.02
(Russia) and by the project 2005SGR00790
(AGAUR, Catalunya, Spain).

\end{document}